\documentstyle[12pt]{article}
\def\be{\begin{equation}}
\def\ee{\end{equation}}

\begin{document}
\thispagestyle{empty}
\setcounter{page}0

{}~\vfill
\begin{center}
{\Large\bf      Search for Tensor Interactions in\\
\ \\
                     Kaon Decays at DA$\Phi$NE
} \vfill

{\large M. V. Chizhov}\footnote{E-mail: $mih@phys.uni$-$sofia.bg$}
 \vspace{1cm}

{\em Bogoliubov Laboratory of Theoretical Physics, Joint Institute for
Nuclear Research, Dubna, Russia}\\ and \\ {\em
Centre of Space Research and Technologies, Faculty of Physics,
University of Sofia, 1126 Sofia, Bulgaria}\footnote{Permanent address}

\end{center}  \vfill

\begin{abstract}

The semileptonic kaon decays $K_{l2\gamma}$ and $K_{l3}$ are
considered. We use the most general forms of the matrix elements for
these decays. Additional terms could arise as a result of new tensor
interactions between quarks and leptons. Such terms have been detected
in a recent experiments.  The high precision experiments at DA$\Phi$NE
may be ideal to confirm these results.

\end{abstract}

\vfill

\newpage

\section{Introduction} \indent

The testing of the Standard Model now reaches its concluding stage. The
most precise data are the ones from the LEP experiments with an
accuracy $\sim1\%$. To describe these data, it is necessary to take
into account the electroweak radiative corrections. The data put strict
constraints on the new physics i.e. new interactions and new particles.
On the other hand the low energy experiments, like particle decays,
also can be used as a good test for the $V$-$A$ interactions.

The recent experiments on semileptonic meson decays
\cite{Bolotov,Akimenko} point to the possible existence of an admixture
of tensor interactions \cite{Pob1,Chiz1}, which can help to reconcile
\cite{Pob2} the results of the previous experiments \cite{Kl2g}.
Unfortunately, the experimental data is poor and the experimental
status of the form factors determination of the three particles
semileptonic decays is obscure \cite{Kl3}. In this connection the
construction of the $\phi$-factory in Frascati
\cite{Daphne} and the $K$-meson decay experiments would help to
check both the Standard Model and the chiral theory \cite{ChPT} with a
good accuracy. The comparatively large energy release in the kaon
decays would allow to analyze the dependence of form factors on the
momentum transfer.

\section{$K_{l2\gamma}$ decay} \indent

  One of the decays, where the existence of tensor interactions
in the weak processes was observed, was the radiative semileptonic
pion decay $\pi^- \to e^- \tilde{\nu} \gamma$ \cite{Bolotov}.
The branching ratio $B^{exp}=(1.61\pm0.23)\times 10^{-7}$
is smaller than the theoretical ratio
$B^{th}=(2.41\pm0.07)\times 10^{-7}$ expected in the
Standard Model. This fact can be explained by the destructive
interference of the standard decay amplitude with
the newly introduced tensor amplitude
\cite{Bolotov,Pob1}
\be
M_T={eG_F \cos\theta_C \over \sqrt 2}~ F_T~ \varepsilon^\alpha q^\beta~
\bar{u}(p_e) \sigma_{\alpha\beta} (1-\gamma^5) v(p_\nu).
\label{1}
\ee
The dimensionless constant $F_T$ describes the strength of the new
interaction relative to the ordinary Fermi coupling. Its experimental
value is $F_T \approx 5\times 10^{-3}$.

At the quark level this amplitude can be described by a four fermion
interaction of the form
\be
{\cal L}^{\Delta S=0}_T = {G_F \cos\theta_C \over \sqrt 2}~ f_T~
\bar{\psi}_u \sigma^{\alpha\beta} \psi_d \cdot
\bar{\psi}_e \sigma_{\alpha\beta} (1-\gamma^5) \psi_\nu + {\rm h.c.}
\label{2}
\ee
The relation between $F_T$ and $f_T$ can be obtained in the
framework of the relativistic quark model \cite{Pob1}
$$
F_T={1 \over 3}~{F_\pi \over m_q}~f_T \approx 0.13~f_T,
$$
where $F_\pi=131$ MeV is the pion decay constant, $m_q=340$ Mev
is the constituent quark mass, or by
applying the QCD techniques and the PCAC hypothesis \cite{Beljaev}
$$
F_T= {2 \over 3}~\chi~{<0|\bar{q}q|0> \over F_\pi}~f_T \approx 0.4~f_T,
$$
here $\chi=-5.7\pm 0.6$ GeV$^{-2}$ is the quark condensate magnetic
susceptibility,
\mbox{$<0|\bar{q}q|0>$}$=-(0.24{\rm GeV})^3$ is the vacuum expectation
value for the quark condensate.

However, the constraints following from the $\pi_{l2}$ decay
\cite{Voloshin}, forbid the existence of tensor interactions (2)
between quarks and leptons
with a coupling constant for the tensor interaction $f_T \sim 10^{-2}$,
which is necessary to explain the experiment \cite{Bolotov}.

In ref. \cite{Gabrieli} an extension of the interaction (2)
was proposed including also the strange quark
\be
{\cal L}^{\Delta S=1}_T = {G_F \sin\theta_C \over \sqrt 2}~ f_T^1~
\bar{\psi}_u \sigma^{\alpha\beta} \psi_s \cdot
\bar{\psi}_l \sigma_{\alpha\beta} (1-\gamma^5) \psi_\nu + {\rm h.c.}
\label{3}
\ee
This interaction can be detected in the kaon decay
$K^- \to l^- \tilde{\nu} \gamma$.
However, it is unreliable to make any
physical predictions \cite{Gabrieli} on the base of the contradictory
situation of the $\pi_{l2}$ decay.

  This problem was solved in ref. \cite{Chiz1},
where a new tensor interaction
with the coupling constant $f_t \sim 0.1$ was introduced
\be
{\cal L}_T=- {G_F \over \sqrt 2}~ f_t~
\bar{\psi}_u \sigma^{\alpha\lambda} \psi_{d(\theta)}~
{Q_{\alpha} Q^{\beta} \over Q^2}~
\bar{\psi}_l \sigma_{\beta\lambda} (1-\gamma^5) \psi_\nu + {\rm h.c.}~,
\label{T}
\ee
dependent on the momentum transfer to the lepton pair $Q$,
where $\psi_{d(\theta)}=\cos\theta_C~\psi_d + \sin\theta_C~\psi_s$.
This interaction appears effectively through the exchange of massive
tensor particles. In ref. \cite{Chiz1} a mechanism
for providing mass of the
tensor particles was proposed, which leads to a pole $Q^2$ in the
interaction (\ref{T}). For particle decays $Q^2$ is always positive and
does not lead to difficulties. In scattering processes
there exists a kinematic region with $Q^2=0$. This could lead to
a diffraction peak, which, however,
had not been observed at the neutrino
scattering experiments \cite{CHARM}. This difficulty can be avoided
if another mass generating mechanism is used for the tensor particles.

The results of ref. \cite{Chiz1} remain valid if the interaction (4)
changes its form
\be
{\cal{L}}_{ql}=- {G_F \over \sqrt 2}~ f_t~
\bar{\psi}_u \sigma^{\alpha\lambda} \psi_{d(\theta)}~
{Q_{\alpha} Q^{\beta} \over m_\pi^2}~
\bar{\psi}_l \sigma_{\beta\lambda} (1-\gamma^5) \psi_\nu + {\rm h.c.}~,
\label{ql}
\ee
by introducing the parameter $m_\pi^2$ which is of the order of the momentum
transfer in the pion or kaon decay. This interaction between quarks and
leptons leads to an additional tensor term $M_T$ in the total
matrix element of the radiative semileptonic kaon decay
$K^+ \to l^+ \nu \gamma$
\be
M = M_{IB} + M_{SD} + M_T,
\label{6}
\ee
where
\be
M_{IB}=i{e G_F \sin{\theta_C} \over \sqrt 2}~ F_K~ m_l~ \varepsilon_{\alpha}\;
\bar{u}(p_\nu) (1+\gamma^5)
\left[ {p^\alpha \over pq} -
{2p_l^\alpha + i \sigma^{\alpha\beta} q_{\beta} \over 2 p_l q} \right] v(p_l)
\label{7}
\ee
is a QED correction (inner bremsstrahlung) to the $K^+ \to l^+ \nu$ decay
with the kaon decay constant $F_K=160$ MeV,
and
\be
M_{SD}=-{eG_F \sin{\theta_C} \over \sqrt 2 m_K} \varepsilon^{\alpha}
\left[ F_V e_{\alpha\beta\rho\sigma}p^{\rho}q^{\sigma} +
iF_A(pq \cdot g_{\alpha\beta}-p_{\alpha}q_{\beta}) \right]
\bar{u}(p_\nu) \gamma^{\beta} (1-\gamma^5) v(p_l)
\label{8}
\ee
is a structure dependent amplitude parametrized by two form factors $F_V$
and $F_A$; $\varepsilon^{\alpha}(q)$ is the photon polarization
vector with $\varepsilon^\alpha q_\alpha=0$,
and
\be
M_T= -{eG_F \sin{\theta_C} \over \sqrt 2 m_K^2}~ F_T
\left\{ Q^2~ \varepsilon^\alpha q^\beta +
[(\varepsilon Q) q^\alpha - (q Q) \varepsilon^\alpha]~Q^\beta\right\}
\bar{u}(p_\nu) (1+\gamma^5) \sigma_{\alpha\beta} v(p_l),
\label{9}
\ee
where $Q = p - q = p_\nu + p_l$ is the momentum transfer
to the lepton pair.

We choose the kinematical variables as the conventional quantities
$x=2pq/m_K^2$ and $y=2pp_l/m_K^2$.
The differential decay width is
\be
{{\rm d}^2 \Gamma \over {\rm d}x {\rm d}y}=
{\alpha \over 2\pi}~{\Gamma_{K\to l\nu} \over (1-r_l)^2}~\rho(x,y).
\label{10}
\ee
Here $r_l\equiv (m_l/m_K)^2$ and the Dalitz plot density $\rho(x,y)$
is given by
\be
\rho(x,y)=\rho_{IB}(x,y) + \rho_{SD}(x,y) + \rho_{IBSD}(x,y)
+ \rho_{T}(x,y) + \rho_{IBT}(x,y) + \rho_{SDT}(x,y),
\label{11}
\ee
where
$$
\rho_{IB}=IB(x,y),~~~
\rho_{SD}(x,y)=a_l^2~\left[(1+\gamma_A)^2~SD^+(x,y)+
(1-\gamma_A)^2~SD^-(x,y)\right],
$$
$$
\rho_{IBSD}=2 a_l\sqrt r_l
\left[(1+\gamma_A)G^+(x,y)+(1-\gamma_A)G^-(x,y)\right],~
\rho_{T}(x,y)=a_l^2 \gamma_T^2~T(x,y),
$$
\be
\rho_{IBT}=2a_l \gamma_T~I(x,y),~~
\rho_{SDT}=2a_l^2 \gamma_T\sqrt r_l
\left[(1+\gamma_A)J^+(x,y)+(1-\gamma_A)J^-(x,y)\right].
\label{12}
\ee
We introduce the following constants
$$
a_l = {m_K^2 \over 2 F_K m_l}~F_V,~~~~~
\gamma_A={F_A \over F_V},~~~~~\gamma_T={F_T \over F_V},
$$
where the form factor $F_V$ can be estimated from the PCAC hypothesis
and the axial anomaly as $F_V=m_K/(4 \pi^2 F_K)$.
The theory does not give exact predictions for
the axial $F_A$ and tensor $F_T$ form factors, which should be
determined experimentally.
The explicit form of the functions $IB(x,y)$, $SD^\pm(x,y)$,
$G^\pm(x,y)$, $T(x,y)$, $I(x,y)$ and $J^\pm(x,y)$ are given in
the appendix.

In the radiative semileptonic pion decay $\pi_{e2\gamma}$ the constant
$a^\pi_e=m_\pi^3 / 8 \pi ^2 F_\pi^2 m_l \approx 4$ and the densities
$\rho_{IB}$ and $\rho_{SD}$ give almost the same contribution to the
decay rate. The densities $\rho_{IBSD}$ and $\rho_{SDT}$ are
negligibly small from chirality considerations. Therefore
the interference term $\rho_{IBT}$ is not suppressed and
gives a considerable contribution to the decay rate. Such kind of a
contribution with a negative sign (destructive interference) has been
observed experimentally \cite{Bolotov}.

In the kaon decay $K^+ \to e^+ \nu_e \gamma$ the constant
$a^K_e \approx 120$ is thirty times as large as
the corresponding value $a^\pi_e$ for the pion decay. This leads to
a considerable decrease in the relative contribution of $\rho_{IB}$
and of the corresponding interference term $\rho_{IBT}$.
In case of the radiative kaon decay to the muon
$K^+ \to \mu^+ \nu_\mu \gamma$ the constant is
$a^K_\mu \approx 0.5$, and the leading contribution
is that of $\rho_{IB}$.
In this decay the interference contribution
$\rho_{IBSD}$ is essential due to the large muon mass, and
$\rho_{IBT}$ is suppressed by the small constant $a^K_\mu$.

We can make the following conclusion: The radiative semileptonic
pion decay $\pi_{e2\gamma}$ is an ideal process where the
tensor interaction reveals with its full strength. In the radiative
semileptonic kaon decays $K_{l2\gamma}$ the tensor interaction effect
is suppressed. Moreover, in the kaon decays there exists a strong
background from the processes $K_{l3}$. Therefore, in order to avoid
the contribution from these processes the experiments are usually made
in the narrow kinematic region $y\approx (1+r_l)$.
In this kinematic region the contribution to the $K_{e2\gamma}$ decay
>from the tensor interactions is negligible, and
actually the experiment is sensitive only to the term $SD^+$.
In case of the $K_{\mu2\gamma}$ decay the contributions from
$SD^+$, $J^+$ and $T$ are considerable, but they have the same
distribution on the variable $x$ at $y=(1+r_\mu)$ and can not
be separated.

There is still another reason for the disability to make more decisive
conclusions about the existence of tensor interactions while analyzing
the radiative kaon decay. As a result of the large energy release, the
resonance exchange with strangeness \cite{Bardin}, can give a
considerable contribution to the vector and axial form factors but
this contribution is only roughly estimated \cite{Ecker}. Therefore, in
order to identify the effects due to tensor couplings, one first has to
pin down the contribution from higher-order effects in chiral
perturbation theory. It is not an easy task to make the estimations
with the required accuracy.

\section{$K_{l3}$ decay} \indent

One of the brightest demonstrations of the existence
of the new interactions is the $K_{l3}$ decay:
$K^+ \to \pi^o l^+ \nu$.
The most general Lorentz invariant form of the matrix element of
this decay is \cite{Steiner}
$$
M={G_F \sin{\theta_C} \over {\sqrt 2}}~ \bar{u}(p_\nu) (1+\gamma^5)
\left\{ m_K F_S - {1 \over 2} \left[ ( P_K + P_{\pi})_{\alpha} f_+
+ (P_K - P_{\pi})_{\alpha} f_- \right] \gamma^{\alpha} \right .
$$
\be
\left . + i{F_T \over m_K} \sigma_{\alpha\beta}
P^{\alpha}_K P^{\beta}_{\pi} \right\} v(p_l).
\ee
It consists of a scalar, a vector and a tensor terms, where the form factors
$F_S$, $f_\pm$ and $F_T$ are functions of the squared momentum
transfer to leptons $Q^2=$\mbox{$(P_K-P_\pi)^2$}. As should be expected
>from the $W$-boson exchange, there exists no compelling evidence
for terms other than those of a pure vector nature. The contribution
of the electroweak radiative corrections to the scalar and tensor
form factors is negligibly small. Therefore, the appearance of a
considerable nonzero scalar and tensor form factors
\cite{Steiner,Akimenko} points to deviations from the standard
$V$-$A$ interaction.

The term in the vector matrix element with $f_-(Q^2)$
can be reduced (using Dirac equation) to the scalar form factor:
$$
(P_K-P_\pi)_\alpha~ \bar{u}(p_\nu) (1+\gamma^5)
\gamma^\alpha v(p_l) =
-m_l~ \bar{u}(p_\nu) (1+\gamma^5) v(p_l).
$$
In the same way the tensor matrix element can be reduced to a vector and
a scalar matrix elements with a distinctive dependence on the
momentum of the latter:
\begin{eqnarray*}
2iP^\alpha_K P^\beta_\pi~ \bar{u}(p_\nu) (1+\gamma^5)
\sigma_{\alpha\beta} v(p_l) &=&
-m_l~ (P_K+P_\pi)_\alpha~ \bar{u}(p_\nu) (1+\gamma^5) \gamma^\alpha v(p_l)
\\
&&
+(P_K+P_\pi)_\alpha~ (p_\nu-p_l)^\alpha~ \bar{u}(p_\nu) (1+\gamma^5) v(p_l).
\end{eqnarray*}
This leads effectively to a redefinition of $f_+$:
$V=f_+ +(m_l/m_K)~ F_T$, \\
and $F_S$:
$S=F_S +(m_l/2m_K)~ f_- +(P_K+P_\pi)_\alpha~ (p_\nu-p_l)^\alpha~
/ (2m_K^2)~ F_T$.

Therefore, the Dalitz plot density in the rest frame of the kaon
\be
\rho(E_\pi,E_l)=A \cdot |V|^2 +
B \cdot {\rm Re}(V^* S) + C \cdot |S|^2
\label{Dal_k3}
\ee
is expressed through $V$ and
$$
S = F_S + {m_l \over 2 m_K}~ f_- +
{1 \over m_K}~\left[(E_\nu-E_l)+{m_l^2 \over 2 m_K}\right]~ F_T.
$$
Here
$$
A=m_K~(2 E_l E_\nu - m_K \Delta E_\pi) -
m_l^2~ (E_\nu - {1 \over 4}~ \Delta E_\pi),
$$
$$
B=m_l~m_K~ (2E_\nu - \Delta E_\pi),~~~~~~~~C=m_K^2~\Delta E_\pi.
$$
where $\Delta E_\pi= E^{max}_\pi - E_\pi$.

One can see that the study of the $K_{\mu 3}$ decay mode alone cannot
give a limit on $F_S$. For the $K_{e3}$ decay, ratio $m_e/m_K \sim
10^{-3}$ is small. The interference terms between the vector and the
scalar matrix elements are also small. This makes it possible to separate
the corresponding contributions from various matrix elements. A
characteristic feature of the tensor matrix element contribution is the
existence of a minimum, in the case when the energies of neutrino $E_\nu$
and the positron $E_e$ are equal.

This is well noticed in the c.m. frame of the lepton pair.
The Dalitz plot density, given in (\ref{Dal_k3}), can be
transformed to the dilepton c.m. system
\be
\rho(E_\pi,\cos\theta) \propto {\varepsilon^5 p_\pi \over E_K}
\left\{ |m_K^2 F_S+\varepsilon p_\pi F_T\cos\theta|^2
+p_\pi^2 |f_+|^2 \sin^2\theta \right\},
\label{cm}
\ee
where $\theta$ is the angle between $\pi^o$ and $e^+$ in the
same system, $p_\pi=\sqrt{E_\pi^2-m_\pi^2}$,
$E_K= \sqrt {p_\pi^2+m_K^2}$ and $\varepsilon=E_K-E_\pi=\sqrt{Q^2}$.

In ref. \cite{Braun} a method for the separation of the form factors
was proposed using integration over the pion spectrum and fitting
to the distribution of $\cos\theta$. However, in case of $F_S$ close
to zero, the tensor contribution dependence on $\cos^2\theta=
1-\sin^2\theta$ could be absorbed into the vector contribution, and
the separation could not be provided in this way. Such a situation
occurs in ref. \cite{Braun}. Meanwhile, we want to note that
our formula (\ref{cm}) for $\rho(E_\pi,\cos\theta)$
differs from the one given in \cite{Braun}.
We want also to point to the lack of consistency in the
definition of the form factors $f_+$ in \cite{Braun}, namely in the
eq. (1) and in the formula for $\Gamma(K_{e3})$ at page 241
(in the last formula it is smaller by $\sqrt 2$ ).

Fitting Dalitz plot to the distribution $\rho(E_\pi,E_e)$
leads to nonzero values of the form factors
$|F_S/f_+(0)|=0.070\pm0.016$ and $|F_T/f_+(0)|=0.53\pm0.10$
\cite{Akimenko}. An analysis was provided in terms
of linearly parametrized
$f_+(Q^2)=f_+(0)~[1+\lambda_+~Q^2/m_\pi^2]$
and constants $F_S$ and $F_T$.
It is important to note here that the separation of the scalar
contribution is the most unreliable, as far as any dependence
$F_T(Q^2)$ on $Q^2$ or deviation from linearity in $f_+(Q^2)$
imitates the existence of a nonzero scalar form factor.

Indeed such a dependence arises, if one relies on the new
tensor interaction between quarks and leptons (\ref{ql}).
In the framework of the relativistic quark model we can get
the following relation between $F_T$ and $f_t$ \cite{Chiz1}
$$
F_T\approx 1.1 {Q^2 \over m_\pi^2}~f_t.
$$
We see that the tensor form factor $F_T$ depends on $Q^2$. Therefore
one has to take into account the momentum transfer dependence of the
form factors when analyzing experimental data.

\section{Conclusion} \indent

DA$\Phi$NE provides an opportunity to improve our knowledge
about the $K$ decays. In this connection we have discussed two
modes of the kaon decays $K_{l2\gamma}$ and $K_{l3}$, where
possible deviation from the Standard Model could be seen.
By kinematical considerations the tensor interactions
can not give a direct contribution to two particles semileptonic
meson decay. Therefore we have analyzed meson decays with
three particles in the final state. The decays with more
particles in the final state lead to a more complicated
kinematics and include additional form factors.

We have shown, that for the $K_{l2\gamma}$ decay the contribution
>from the new tensor interaction is suppressed. Moreover, the
strong background from the $K_{l3}$ decay reduces the kinematic region
available for analysis. Fortunately, the $K_{l3}$ decay is one,
where the presence of the new interactions can be detected.
Analyzing the Dalitz plot distribution for the $K^+ \to \pi^o e^+ \nu$
decay we can very well separate the contributions from the vector
and tensor terms. The presence of a nonzero tensor or scalar form
factor would be a signal of the existence of new interactions in
the weak processes.

\vspace{2cm}

{\bf Acknowledgments} \vspace{4mm}

The author would like to thank L.V.Avdeev and D.P.Kirilova
for the careful reading of the manuscript and the helpful remarks.
He also acknowledges the hospitality of the Bogoliubov
Laboratory of Theoretical Physics, JINR, Dubna,
where this work has been completed.

\pagebreak[4]
{\bf Appendix} \vspace{4mm}

In this appendix we give an analytical expressions for the functions
$IB(x,y)$, $SD^\pm(x,y)$, $G^\pm(x,y)$, $T(x,y)$, $I(x,y)$ and $J^\pm(x,y)$:
\begin{eqnarray*}
&&IB(x,y) = {1-y+r_l\over x^2(x+y-1-r_l)}
\left[x^2+2(1-x)(1-r_l)-{2x r_l(1-r_l)\over x+y-1-r_l}\right],
\\
\\
&&SD^+(x,y) = (x+y-1-r_l) \left[(x+y-1)(1-x)-r_l\right],
\\
\\
&&SD^-(x,y) = (1-y+r_l) \left[(1-x)(1-y)+r_l\right],
\\
\\
&&G^+(x,y) = {1-y+r_l\over x(x+y-1-r_l)}
\left[(x+y-1)(1-x)-r_l\right],
\\
\\
&&G^-(x,y) = {1-y+r_l\over x(x+y-1-r_l)}
\left[(1-x)(1-y)-x+r_l\right],
\\
\\
&&T(x,y) = 2(1-x)^2 (1-y)(x+y-1)+r_l(1-x)(2-x)(x+2y-2)
\\
\\
&&~~~~~~~~~~-r_l^2\left[x^2+2(1-x)\right],
\\
\\
&&I(x,y) = -(1-x)(1-y+r_l),
\\
\\
&&J^+(x,y) = x \left[(x+y-1)(1-x)-r_l\right],
\\
\\
&&J^-(x,y) = x(1-x)(1-y+r_l).
\end{eqnarray*}

\pagebreak[4]

\end{document}